\documentstyle[a4,12pt]{article}
\begin{document}

\begin{titlepage}
\hfill{hep-th/9404069}

\vspace*{\fill}

\centerline{\huge Conservation laws for}
\centerline{\ }
\centerline{\huge the classical Toda field theories.}
\vspace{2cm}

\centerline{\large Erling G.~B.~Hohler and K\aa re Olaussen}
\vspace{0.2cm}
\centerline{\large Institutt for fysikk, NTH,}
\vspace{0.1cm}
\centerline{\large Universitetet i Trondheim}
\vspace{0.1cm}
\centerline{\large N--7034 Trondheim, Norway.}

\vspace{0.5cm}

\centerline{July 9, 1993}

\vspace{2.5cm}

\begin{abstract}
We have performed some explicit calculations
of the conservation laws for classical (affine) Toda
field theories, and some generalizations
of these models. We show that there is a huge
class of generalized models which have an infinite
set of conservation laws, with their integrated
charges being in involution.
Amongst these models we find that only the $A_m$ and $A_m^{(1)}$
($m\ge 2$) Toda field theories admit such conservation laws for spin-3.
We report on our explicit calculations
of spin-4 and spin-5 conservation laws in the
(affine) Toda models.
Our perhaps most interesting
finding is that there exist conservation
laws in the $A_m$ models ($m\ge4)$ which have a different
origin than the exponents of the corresponding affine
theory or the energy-momentum tensor of a conformal
theory.
\end{abstract}

\vspace*{\fill}

\end{titlepage}

\section{Introduction}

There is an intimate connection between the integrability of
a Hamiltonian system and the existence of a sufficient number
of conserved quantities.
Liouville proved that if a system with $N$ degrees of freedom (i.e.,
with a $2N$-dimensional phase space) has $N$ independent first
integrals in involution (i.e.\ with mutually vanishing Poisson brackets),
then the system is integrable by quadratures\cite{Arnold}.
Conversely, an integrable system with $N$ degrees of freedom has $N$
independent conservation laws in involution.
During the last quarter of a century there has been great
progress in the understanding of integrable systems with {\em infinitely}
many degrees of freedom, e.g.\ certain 1+1 dimensional non-linear partial
differential equations\cite{Miura}.
Such systems must necessarily have an infinite number of conserved
quantities in involution. Since they are governed by
local evolution equations the corresponding conservation laws can
be written in a local form
\begin{equation}
    {\overline\partial} {J}_S + {\partial}{\overline J}_S = 0,
    \ \ \ \mbox{$S=1,2,\ldots,\infty$}
    \label{conslaw}
\end{equation}
where $\partial \equiv {\partial}/{\partial t} + {\partial}/{\partial x}$
and ${\overline\partial} \equiv {\partial}/{\partial t}-{\partial}/{\partial
x}$.
Since there are many ways to count to infinity it is a more delicate question
to
decide when the existence of an infinite number of independent conservation
laws imply complete
integrability.

We here report on some investigations of the conservation laws for models
defined by Lagrangians of the type
\begin{equation}
   {\cal L} = \frac{1}{2}\,\partial\varphi\!\cdot\!{\overline\partial}\varphi -
   V(\varphi) \equiv
   \frac{1}{2}\,\partial\varphi\!\cdot\!{\overline\partial}\varphi -
              \sum_{r=1}^n\; v_r(\varphi), \label{Toda}
\end{equation}
where $\varphi$ is a real $m$-component field, and
\(
   v_r(\varphi) =  a_r\,\exp\left( \alpha_r \!\cdot\! \varphi\right)
\).
Here the $\alpha_r$'s are distinct real $m$-component vectors, and the $a_r$'s
are real numbers.
To this class of models belong the so-called
Toda field theories, exemplified by the potential (in appropriate dimensionless
units)
\begin{equation}
   V(\varphi) = \sum_{r=1}^{m-1}\; \exp\left( \varphi_{r} - \varphi_{r+1}
\right)
   \equiv W_{m}, \label{TFT}
\end{equation}
and the affine Toda field theories, exemplified by the potential
\begin{equation}
   V(\varphi) = W_{m} + \exp\left( \varphi_{m} - \varphi_{1} \right),
\label{ATFT}
\end{equation}
i.e.\ the model (\ref{TFT}) with periodic boundary conditions.
We note that with an $x$-independent field $\varphi$ these models reduce to
respectively
the open and periodic Toda lattices\cite{Toda}, hence their names.

For a simple physical interpretation these models can be viewed as a collection
of (infinitely long) parallell strings, with a (somewhat strange) interaction
potential between adjacent strings.
Another possible physical interpretation is as a collection of
parallell wires, with exponential inductance for the leakage currents
between adjacent wires. The fields then represent the voltages on the wires.
However, the main interest in these models stems from the fact that they after
quantization may represent a huge class of conformal field theories,
or integrable perturbations of such theories. Thus, they have the potential
of describing the relevant degrees of freedom of many 2-dimensional statistical
models at---and close to---their critical points. They also provide interesting
examples of interacting relativistic quantum field theories.

In a general Toda field theory the $\alpha_r$'s constitute the simple positive
roots
of a Lie algebra, and in a general affine Toda field theory they constitute
these roots augmented with an additional vector which is some linear
combination
of the roots of the Lie algebra\footnote{In many, but not all, cases this
vector is
the negative of the maximal root of the Lie algebra.}.
In mathematicians nomenclature
the examples (\ref{TFT}, \ref{ATFT}) correspond to the algebras $A_{m-1} \equiv
SU(m)$
and $A_{m-1}^{(1)}$. Readers who are unfamiliar with the language of Lie
algebras
should not feel intimidated by these statements,---they are just a peculiar
(but concise)
way of decribing some of the possible potentials $V(\varphi)$.
Some explicit examples of model potentials with their
associated algebras are listed in Table 1.
\begin{table}
{\small
\begin{center}

\begin{tabular}{|c|c||c|c|}
\hline
$V(\varphi)$ & Algebra & $V(\varphi)$ & Algebra \\
\hline
$W_m + \exp\left( \varphi_m \right)$ & $B_m$ &
$W_3 + \exp\left(\varphi_2+\varphi_3-2\,\varphi_1 \right)$ & $D_4^{(3)}$\\
$W_m + \exp\left( 2\,\varphi_m\right)$ & $C_m$  &
$W_m + \exp\left(\varphi_m \right) + \exp\left( -\varphi_1-\varphi_2\right)$ &
$B_m^{(1)}$\\
$W_m + \exp\left(\varphi_{m-1}+\varphi_{m}\right)$ & $D_m$ &
$W_m + \exp\left(2\,\varphi_m \right) + \exp\left(-2\,\varphi_m \right)$&
$C_m^{(1)}$\\
$W_m + \exp\left(\varphi_m-\varphi_1 \right)$ & $A_{m-1}^{(1)}$ &
$W_{2m} + \exp\left(2\,\varphi_{2m}\right) + \exp\left(-\varphi_1\right)$ &
$A_{2m}^{(2)}$\\
$W_2 + \exp\left(\frac{1}{2}\varphi_2-\frac{1}{2}\varphi_1\right)$ &
$A_2^{(2)}$ &
$W_{2m-1} +
\exp\left(2\,\varphi_{2m-1}\right)+\exp\left(-\varphi_1-\varphi_2\right)$ &
$A_{2m-1}^{(2)}$\\
$W_m + \exp\left(\varphi_m\right)+\exp\left(-\varphi_1\right)$ & $D_m^{(2)}$ &
$W_m + \exp\left(\varphi_{m-1} + \varphi_{m} \right) +
\exp\left(-\varphi_1-\varphi_2\right)$ &
$D_m^{(1)}$\\
\hline
\end{tabular}

\end{center}
}
\caption{Some potentials of (affine) Toda field theories, with the
nomenclature for their associated algebras. There are no universally accepted
names
for the affine Lie algebras, we are following the conventions of
Kac\protect{\cite{Kac}}.
Here $m\ge2$ for $A_{2m}^{(2)}$ and $A_{2m-1}^{(2)}$. To avoid equivalences one
may also
make the restrictions $m\ge2$ for $B_m$, $m\ge3$ for $C_m$, and $m\ge4$ for
$D_m$.}

\end{table}

By simply looking at the potentials in Table~1 it is rather difficult to see
any
direct significance of the Lie algebras for the models they specify. Although
the
existence of these algebras have been instrumental in proving integrability of
the (classical) models\cite{Leznov}, and the existence of an infinite set of
(classical)
conservation laws in involution\cite{Olive}, this does not rule out that there
may exist
additional models in the class (\ref{Toda}) which are either integrable or
at least have an infinite set of local conservation laws.
We thus set out to investigate the conservation laws of models (\ref{Toda})
in more generality, and indeed found additional examples which---as a
consequence of conformal
symmetry---have an infinite set of (classical) conservation laws.

\section{Field equations and spin-1 conservation laws}

We now turn to the Lagrangian (\ref{Toda}). The components of the vectors
$\alpha_r$ may be collected into a $n \times m$ matrix
\(
   A_{ri} \equiv (\alpha_r)_i
\).
Not all such matrices represent truly different models, since we may make
orthogonal transformations on the field components,
$\varphi_i = O_{ij}\, \varphi'_j $, and permute the order of the potentials
$v_r$.
I.e., models decribed by the matrices $A_{ri}$ and
(summation convention is used unless noted otherwise)
\[
  A'_{ri} = P_{rs}\, A_{sj}\,O_{ji},
\]
where $P_{rs}$ is a permutation matrix and $O_{ji}$ is an orthogonal matrix,
are
essentially equivalent. Such transformations with
$A_{ri} = A'_{ri}$ are symmetries of the model.
Further, depending on the solvability of the equations
\[
    A_{ri}\,\Delta\varphi_i + \log\left(\vert L^2\,a_r \vert\right) = 0,
\]
the coefficients in front of the exponentials may be transformed to a simpler
form by a constant shift of the fields $\varphi\to\varphi+\Delta\varphi$
(while remaining proportional to a dimensional quantity $L^{-2}$, where $L$ is
some length).

The equations of motion become
\begin{equation}
   -{\overline\partial}{\partial}\,\varphi_i = v_r\,A_{ri}. \label{motion}
\end{equation}
To each right eigenvector $\chi^{(k)}$ of $A$ with zero eigenvalue,
the field $\varphi^{(k)} = \varphi\cdot\chi^{(k)}$ satisfies a free field
equation,
\(
{\overline\partial}{\partial}\,\varphi^{(k)}=0
\).
The corresponding spin-1 currents ${\cal J} = \partial\varphi^{(k)}$ and
${\overline{\cal K}} = {\overline\partial}\varphi^{(k)}$ are conserved,
\begin{equation}
   {\overline\partial}{\cal J} = 0,\ \ \ {\partial}{\overline{\cal K}} = 0.
\label{1conslaw}
\end{equation}
By dimensional analysis and (1+1-dimensional) Lorentz invariance it follows
that
a general spin-$S$ conservation law is expressible in the form (\ref{conslaw}),
with
\begin{equation}
  J_S = {\cal P}(\partial\varphi,\partial^2\varphi,\ldots),\ \ \
  {\overline J}_S = v_r\,{\cal Q}_r(\partial\varphi,\partial^2\varphi,\ldots),
  \label{sconslaw}
\end{equation}
or the parity transformation of this (i.e.\ $\partial \rightleftharpoons
{\overline\partial}$,
$J_S \to {\overline K}_S$, ${\overline J}_S \to K_S$). In equation
(\ref{sconslaw}) ${\cal P}$
is a polynomial of dimension $S$, and ${\cal Q}_r$ is a polynomial of dimension
$S-2$. Here
$\partial$ and $\overline\partial$ have (inverse length) dimensions 1,
$\varphi$ has dimension $0$,
and $v_r$ has dimension 2. Then, by ordinary rules of differentiation,
equations (\ref{1conslaw})
automatically imply two classes of spin-$S$ conservation laws,
\begin{equation}
   {\overline\partial} {\cal P}({\cal J},\partial{\cal J},\ldots)=0,\ \ \
   {\partial}{\cal P}({\overline{\cal K}},{\overline\partial}{\overline{\cal
K}},\ldots)=0.
\end{equation}
However, these are rather obvious consequences of the free fields present in
the model.
To maintain focus on the more interesting conservation laws we shall
assume that all free fields have been separated from the model, so that the
matrix $A_{ri}$
have no right eigenvectors with zero eigenvalue. This means in particular that
$n \ge m$, and
that all $m \times m$ submatrices of $A_{ri}$ are invertable.

\section{Conservation laws due to conformal invariance}

In this section we search for spin-2 conservation laws, essentially the
conservation
of the energy momentum tensor. We are in particular searching for those models
where
the conservation can be written in a form similar to (\ref{1conslaw}),
\[
  {\overline\partial}T=0,\ \ \ \ {\partial}{\overline T}=0,
\]
as is typical for the energy-momentum tensor of a conformal field theory.
Again, by ordinary
rules of differentiation this will imply the existence of an infinite set of
higher spin
conserved currents,
\[
        {\overline\partial}{\cal P}(T,\partial T,\ldots) = 0,\ \ \ \
        {\partial}{\cal P}({\overline T},\partial{\overline T},\ldots) = 0,
\]
where $\cal P$ is a polynomial of dimension $S$, with $T$, ${\overline T}$
being
of dimension 2.
The general form of the spin-2 currents can be written as
\begin{equation}
    J_2 = \beta_{ij}\,\partial\varphi_i\,\partial\varphi_j +
\rho_i\,\partial^2\varphi_i,\ \ \
    {\overline J}_2 = v_r\,C_r,
\end{equation}
(or their parity transforms) where $\beta_{ij}$ is symmetric. Inserting this
ansatz into
equation~(\ref{conslaw}) and using the equations of motion (\ref{motion}) we
find that
the equation (no sum over $r$)
\[
   C_r\,A_{rj} = 2\,A_{ri}\,\beta_{ij} + A_{ri}\,\rho_i\,A_{rj}
\]
must hold. With $\beta_{ij}= \frac{1}{2}\delta_{ij}$ we always have the
solution
$C_r = 1 + A_{ri}\,\rho_i$. With $\rho_i=0$ this leads to the canonical
energy-momentum
tensor, as i.e.\ calculated from Noether's theorem. Here we instead search for
models
where $C_r=0$ is a possible solution, since these will be conformally
invariant.
Let $u$ denote the $n$-component vector with all entries equal to $-1$. Then
the equation
to be solved is \( A_{ri}\,\rho_i = u_r \). The criterium for solvability is
seemingly
modest: that all left eigenvectors $\xi^{(k)}$ of $A$ with eigenvalue 0 must be
orthogonal to $u$. However, this is enough to rule out all the affine Toda
theories.
To investigate which matrices $A$ lead to conformal models, assume first that
$n=m$.
Then $A$ is invertible by assumption, and a solution $\rho$ always exist.
Next we may extend $A$ to an arbitrary $n\times m$, matrix by adding
interaction
potententials $v_r=a_r\,\exp\left(\alpha_r\cdot\varphi\right)$,
$r=m+1,\ldots,n$ with $\alpha_r=\gamma_r - \rho/\rho^2$, where
the $\gamma_r$'s are distinct vectors such that $\gamma_r\cdot\rho=0$. In fact,
it seems that all potentials
\begin{equation}
   V(\varphi) = \exp\left(-\rho\cdot\varphi/\rho^2 \right)\,U(\varphi_{\perp}),
\end{equation}
with $U$ arbitrary, lead to a conformal model. Here $\varphi_{\perp} =
\varphi-\left(\rho\cdot\varphi/\rho^2 \right)\,\rho$. The conserved current
is a light-cone component of the (traceless) energy-momentum tensor,
\begin{equation}
  T = \frac{1}{2}\,\partial\varphi\!\cdot\!\partial\varphi +
\rho\!\cdot\!\partial^2\varphi
  \label{Ttensor}
\end{equation}
(or its parity transformation $\overline T$).

To investigate whether the conservation laws in the above conformal models
are in involution, we have to turn to a canonical formulation. Contrary to
our notation we prefer to use ordinary space-time ($x$,$t$) variables. I.e,\
our basic canonical variables are the fields $\varphi(x,t)$ and
$\pi(x,t)=\partial\varphi(x,t)/\partial t$, with equal time Poisson brackets
$\left\{\pi_i(x,t),\varphi_j(y,t) \right\} = \delta_{ij}\,\delta(x-y)$.
Expressions
like $\partial^n\varphi$ should thus be interpreted as shorthand notation for
a perhaps lengthy expression in the fields $\varphi$, $\pi$, and their spatial
derivatives. This expression can be obtained by eliminating all time
derivatives
of fields with the use of the equations of motion. It is convenient to
introduce
the combinations
\begin{equation}
   \psi_i(x) = \partial\varphi_i(x) = \pi_i(x)+\varphi'_i(x),\ \ \
   {\overline\psi}_i(x) = {\overline\partial}\varphi_i(x) =
\pi_i(x)-\varphi'_i(x). \label{psis}
\end{equation}
Here and in the following we drop explicit references to time dependence. All
fields
are assumed to be evaluated at the same fixed time. A ${}'$ denotes
differentiation
with respect to the remainding (space) coordinate.
The following analysis is much inspired by some remarks by
Freeman and West\cite{West}.
The fields (\ref{psis}) have Poisson
brackets
\begin{equation}
   \left\{ \psi_{i}(x),\psi_j(y) \right\} =
  -\left\{{\overline\psi}_i(x), {\overline\psi}_j(y) \right\} =
  -\delta_{ij}\,\delta'(x-y),\ \ \
   \left\{ \psi_i(x),{\overline\psi}_j(y)\right\} = 0,
\end{equation}
and we further find that
\(
  \left\{\psi_i(x), V(\varphi(y)) \right\} = \left\{ {\overline\psi}_i(x),
V(\varphi(y)) \right\}
  = -f_i(x)\,\delta(x-y)
\),
where $f_i= v_r\,A_{ri}$. Now we rewrite the expression (\ref{Ttensor}) for $T$
in terms of
canonical fields, and evaluate the Poisson bracket
\begin{equation}
  \left\{ T(x), T(y) \right\} = 2\,\left[ T(x)\,\partial_x +
\partial_x\,T(x) - 4\rho^2\,\partial_x^3 \right]\,\delta(x-y). \label{KdV}
\end{equation}
There is an identical expression with $T\to {\overline T}$, while the Poisson
bracket between
$T$ and $\overline T$ is zero. Equation (\ref{KdV}) is the spatial version of
the Virasoro
algebra, but it is also a Poisson bracket for the KdV hiearchy of equations.
I.e.\ with the
bracket (\ref{KdV}) and $H=\int\,dx\, T(x)^2$ as the Hamiltonian, the equation
$T_t= \left\{T,H\right\}$ becomes the KdV-equation. Since this equation is
known to have an
infinite set of conserved quantities in involution, see ref.~\cite{KdV4},
the first two of which are $\int\,dx\,T(x)$ and
$\int\,dx\,T(x)^2$, the same will necessarily be true when the same quantities
are viewed
as coming from a conformal field theory.

To rewrite (\ref{KdV}) in the conventional Virasoro form, we assume the
$x$-coordinate to
be periodic with period $\ell$, and introduce the Fourier modes
\begin{equation}
    L_m = 2\pi\,\rho^2\delta_{m0} +
   \frac{\ell}{4\pi}\,\int_0^{\ell}\,dx\;
   \exp\left(\frac{2\pi i m x}{\ell} \right)\,T(x).
\end{equation}
Then equation (\ref{KdV}) implies the Virasoro algebra
\begin{equation}
   i\left\{ L_m, L_n \right\} = \left(m-n\right)\,L_{m+n} +
   4\pi\,\rho^2 m\left(m^2-1\right)\,\delta_{m+n,0},
\end{equation}
which shows that the model has central charge $c=48\pi\,\rho^2$.

\section{Spin-3 conservation laws}

We now make a general search for spin-3 conservation laws. Since we may always
shift the current by the gradient of an antisymmetric tensor,
\(
    J \to J + \partial X,\ \ \ {\overline J} \to {\overline J} -
{\overline\partial} X,
\)
it suffices to search for expressions of the form
\begin{equation}
   J_3 = {\alpha}_{ijk}\,\partial\varphi_i\, \partial\varphi_j\,
\partial\varphi_k
   + {\beta}_{ij} \partial\varphi_i\,\partial^2\varphi_j,\ \ \
   {\overline J}_3 = v_r\,B_{ri}\,\partial\varphi_i
\end{equation}
(or their parity transforms). Here $\alpha_{ijk}$ is completely symmetric, and
$\beta_{ij}$ is antisymmetric. By performing the differentiations
and using the equations of motion (\ref{motion}) we find that
equation~(\ref{conslaw}) implies
that
\(
  B_{ri} = A_{rk}\,\beta_{ki}
\),
and in turn (no summation convention)
\begin{equation}
  3\sum_{k=1}^m\; A_{rk}\,\alpha_{kij} = \sum_{k=1}^m\; \left(
A_{ri}\,A_{rk}\,\beta_{kj}
   + A_{rj}\,A_{rk}\,\beta_{ki} \right).  \label{alpha}
\end{equation}
To proceed we now restrict the index $r$ to an arbitrary $m$-component subset
of
$\left\{ 1,\ldots,n \right\}$. The matrix $A_{ri}$ is
invertible in the corresponding subspace. Multiplying from the left by this
inverse
matrix $A^{-1}$,
utilizing the complete symmetry of $\alpha_{kij}$, and finally
multiplying the resulting equation by $A_{sk}\, A_{ti}\, A_{uj} $ and summing
over $kij$, we obtain (no summation convention)
\begin{equation}
   K_{st}\, {\tilde\beta}_{su} + K_{su}\,{\tilde\beta}_{st} =
   K_{ut}\,{\tilde\beta}_{us} + K_{us}\,{\tilde\beta}_{ut}, \label{master}
\end{equation}
where $K = A\,A^{T}$ and
${\tilde\beta} = A\,\beta\,A^T$,
with $A^T$ denoting the transpose of $A$. This equation must hold
for the full set of indices $\left\{1,\ldots n\right\}$, since it
is supposed to hold for arbitrary $m$-component subsets.
Note that (\ref{master}) is invariant under
orthogonal transformations of the fields, $\varphi_i = O_{ij}\,{\varphi'_j} $.
When two indices (say $t$ and $u$) are
equal equation (\ref{master}) simplifies to (no summation convention)
\begin{equation}
   \left( K_{tt} + 2\,K_{st}\right)\,{\tilde\beta}_{st}=0.
\end{equation}
{}From this equation, and the (anti)symmetry of the matrices,
we immediately see that $\beta_{st}\ne0$ imply $K_{ss}=K_{tt}=
-2\,K_{st}=-2\,K_{ts}$, and if also $\beta_{tu}\ne0$ then in addition
$K_{uu}=K_{ss}=
-2\,K_{tu}=-2\,K_{ut}$. Equipped with this information one may start
analyzing the general situation by considering (\ref{master})
for all combinations of indices $stu$ within larger
and larger subsets (which by a permutation may be mapped into
$\left\{1,\ldots,\ell\right\}$). Modulo permutation of indices and
multiplication by
constants we only find the solutions
\begin{equation}
  K_{rs} = \left( 2\,\delta_{r,s} - \delta_{r+1,s} - \delta_{r-1,s} \right),\ \
\
  {\tilde\beta}_{rs} = \left( \delta_{r+1,s} - \delta_{r-1,s} \right),
  \ \ \mbox{with $r,s$ in $\left\{1,\ldots n\right\}$.}
\end{equation}
Here the $\delta_{rs}$'s should either be interpreted to vanish when the
indices $r,s$
are outside the range $\left\{1,\ldots,n\right\}$, in which case
$n=m$, $m\ge 2$ and the model is equivalent
to the $A_{m}$ Toda field theory (\ref{TFT}),
or to be periodically extended (so that
index values $n+k$ and $k$ are equivalent),
in which case $n=m+1$ and the model is equivalent\footnote{In these two cases
the matrices $A_{ri}$ allows one to transform all the coefficients $a_r$ (in
$v_r(\varphi)$) to
the same absolute value. In principle there is still an arbitrary choice of
sign for
each of them. However, all $a_r$ must be positive for the potential
$V(\varphi)$
to be bounded from below.}
to the $A^{(1)}_{m}$
affine Toda field theory (\ref{ATFT}). In both cases the model have fields
$\varphi_1,\ldots,\varphi_{m+1}$ obeying the constraint
$\sum_{i=1}^{m+1}\;\varphi_i = 0$.
It is already known from the work of Olive and
Turok\cite{Olive} that the $A_n^{(1)}$-models ($n\ge2$) are the only affine
Toda field
theories with spin-3 conservation laws. We have shown that there is
no other such models (apart from the expected $A_n$ ones) within the larger
class
(\ref{Toda}).

To find explicit expressions for the conserved currents it is convenient to
introduce new fields,
\(
     \theta_r \equiv                \sum_{i=1}^r    \; \varphi_i
\).
These are defined such that
\(
      \theta_{r}\,A_{ri} = \theta_{i} - \theta_{i-1} = \varphi_i
\),
which imply
\[
     \partial\varphi_i\,\beta_{ij}\,\partial^2\varphi_j =
     \partial\theta_r\,{\tilde\beta}_{rs}\,\partial^2\theta_s,\ \
     v_r\,B_{ri}\,\partial\varphi_i =
v_r\,{\tilde\beta}_{rs}\,\partial\theta_s,
\]
and, with the use of (\ref{alpha}),
\[
   \sum_{ijk}\,\alpha_{ijk}\,{\partial\varphi}_i\,
   {\partial\varphi}_j\,{\partial\varphi}_k
   = \frac{1}{3}\,\sum_{rst}\;\left(K_{rs}\,{\tilde\beta}_{rt} +
   K_{rt}\,{\tilde\beta}_{rs} \right) {\partial\theta}_r\,
   {\partial\theta}_s\,{\partial\theta}_t.
\]
Inserting the explicit expressions for $K$ and $\tilde\beta$ we finally obtain
\begin{equation}
  J_3 =
\sum_{r=1}^m\;\left[2\,\left(\partial\theta_r\right)^2\,
\left(\partial\theta_{r+1}
-\partial\theta_{r-1} \right) +
\left(\partial\theta_r\right)\,\left(\partial^2\theta_{r+1}
-\partial^2\theta_{r-1}\right)\right],
\end{equation}
with $\theta_0=\theta_{m+1}=0$.
This expression looks identical for the $A_m$ and $A_m^{(1)}$ models. However,
when rewritten
in terms of canonical variables $\varphi(x)$, $\pi(x)$ they become
different---since the
fields obey different equations of motion in the two cases.
Further
\begin{equation}
  {\overline J}_3 = \sum_{r=1}^n\; v_r\,\left( \partial\theta_{r+1} -
\partial\theta_{r-1} \right),
\end{equation}
where the indices should be interpreted cyclically modulo $m+1$. This
expression is different
for the $A_m$ and $A_m^{(1)}$ models, due to the additional $(m+1)$'st term in
the latter case.

\section{Spin-4 and spin-5 conservation laws}

We have repeated the analysis of the preceeding section for the spin-4 and
spin-5
conservation laws, so far under less general assumptions. In the spin-4 case
one
may write
\[
     J_4 = \beta_{ij}\,\partial^2\varphi_i\,\partial^2\varphi_j + \cdots,
\]
with $\beta$ symmetric, and in the spin-5 case one may write
\[
     J_5 = \beta_{ij}\,\partial^2\varphi_i\,\partial^3\varphi_j + \cdots,
\]
with $\beta$ antisymmetric. Similar ansatzes for ${\overline J}_4$ and
${\overline J}_5$ can
be written down. There are many possible terms in each expressions, also when
all
opportunities of subtracting gradients of antisymmetric tensors have been used.
Inserting
the general expressions for $J$ and $\overline J$ into (\ref{conslaw}) and
using (\ref{motion})
one obtains equations involving the different coefficients. These may be solved
successively
until one is left with a set of equations involving components of the
matrices ${\tilde\beta}$ and $K$. When a solution matrix $\tilde\beta$ is found
the
complete expressions for the conserved currents can be generated by back
substitution. Thus, the problem of finding conservation laws reduces to finding
solution matrices $\tilde\beta$.

In analysing these equations we have
(so far) assumed that the $K$'s come from a Toda field theory or an
affine Toda field theory. For the affine models Olive and Turok\cite{Olive}
found
that there should exist (non-trivial) conservation laws for all spins $S$
such that $S-1 = m \bmod h$, where $m$ is an exponent of the algerbra and $h$
is its Coxeter number. We have explicitly calculated the spin-4 and spin-5
conservation laws for all affine Lie algebras, relying heavily on
algebraic manipulation programs\cite{mathematica}. For the cases treated by
Olive and Turok we find no additional conservation laws. They did however not
treat the $D_{2n}^{(1)}$ theories, which have some exponents occuring with
multiplicity 2. In particular, this is true for exponent 3
of $D_4^{(1)}$. One may suspect that the corresponding model has {\em two}
independent spin-4 conservation laws. This indeed turns out to be the case.
Thus, there seems to be a conservation law for each exponent, counting
multiplicities.

The conservation laws for the affine Toda theories carry over to the Toda
theories.
However, we have found that there occur additional conservation laws
in the latter models. For the spin-4 case this can be interpreted as a
consequence
of conformal symmetry. In this case the additional law can be written as
\(
            {\overline\partial}\,T^2 = 0,
\)
where $T$ is (a part of) the energy-momentum tensor. For all Lie algebras
except $A_1$
the conserved densities of the form
\begin{equation}
      {\overline\partial}\,{\cal P}(T,\partial T,\ldots) = 0,
\label{conform}
\end{equation}
are different from those that correspond to the exponents of the affine Lie
algebras.
But we have also found an additional conservation law in the spin-5 case for
the
$A_m$ models when $m\ge4$. This one cannot be accounted for by expressions
like $(\ref{conform})$. Thus, there exist Toda field theories with conservation
laws
of a different origin than the exponents of the corresponding affine Lie
algebra or
the energy-momentum tensor of a conformal theory.

More details of our results and calculations will be presented elsewhere. The
published version of this letter has appeared in \cite{EH}

\end{document}